\documentstyle[prl,aps,multicol,epsfig]{revtex}

\input epsf.tex

\newcommand{\be}{\begin{equation}}
\newcommand{\ee}{\end{equation}}
\newcommand{\ba}{\begin{eqnarray}}
\newcommand{\ea}{\end{eqnarray}}
\newcommand{\ban}{\begin{eqnarray*}}
\newcommand{\ean}{\end{eqnarray*}}

\newcommand{\braket}[2]{\mbox{$ \langle #1 | #2 \rangle $}}

\newcommand{\ket}[1]{\mbox{$ | #1 \rangle $}}
\newcommand{\bra}[1]{\mbox{$ \langle #1 | $}}

\newcommand{\si}{\sigma}
\newcommand{\demi}{\frac{1}{2}}

\newcommand{\one}{\leavevmode\hbox{\small1\normalsize\kern-.33em1}}

\begin{document}

\title{Quantum cryptography protocols robust against
photon number splitting attacks for weak laser pulses
implementations}
\author{Valerio Scarani$^{1}$, Antonio Ac\'{\i}n$^{1}$, Gr\'egoire Ribordy$^{2}$ and Nicolas Gisin$^{1}$}
\address{
$^{1}$ Group of Applied Physics, University of Geneva, 20, rue de
l'Ecole-de-M\'edecine, CH-1211 Geneva 4, Switzerland\\
$^{2}$ Id-Quantique, rue Cingria 10, CH-1205 Geneva, Switzerland}
\date{\today}
\maketitle

\begin{abstract}
We introduce a new class of quantum quantum key distribution
protocols, tailored to be robust against photon number splitting
(PNS) attacks. We study one of these protocols, which differs from
the BB84 only in the classical sifting procedure. This protocol is
provably better than BB84 against PNS attacks at zero error.
\end{abstract}

\begin{multicols}{2}

Quantum cryptography, or more precisely quantum key distribution
(QKD) is the only physically secure method for the distribution of
a secret key between two distant partners, Alice and Bob
\cite{review}. Its security comes from the well-known fact that
the measurement of an unknown quantum state modifies the state
itself: thus an eavesdopper on the quantum channel, Eve, cannot
get information on the key without introducing errors in the
correlations between Alice and Bob. In equivalent terms, QKD is
secure because of the no-cloning theorem of quantum mechanics: Eve
cannot duplicate the signal and forward a perfect copy to Bob.

In the last years, several long-distance implementations of QKD
have been developed, that use photons as information carriers and
optical fibers as quantum channels \cite{review}. Most often,
although not always \cite{grgr}, Alice sends to Bob a {\em weak
laser pulse} in which she has encoded the bit. Each pulse is {\em
a priori} in a coherent state $\ket{\sqrt{\mu} e^{i\theta}}$ of
weak intensity, typically  $\mu\approx 0.1$ photons. However,
since no reference phase is available outside Alice's office, Bob
and Eve have no information on $\theta$. Consequently, they see
the mixed state $\rho=\int\frac{d\theta}{2\pi}\ket{\sqrt{\mu}
e^{i\theta}}\bra{\sqrt{\mu} e^{i\theta}}$. This state can be
re-written as a mixture of Fock states,
$\sum_{n}p_n\ket{n}\bra{n}$, with the number $n$ of photons
distributed according to the Poissonian statistics of mean $\mu$,
$p_n=p_n(\mu)=e^{-\mu}\mu^n/n!$. Because two realizations of the
same density matrix are indistinguishable, QKD with weak pulses
can be re-interpreted as follows: Alice encodes her bit in one
photon with frequency $p_1$, in two photons with frequency $p_2$,
and so on, and does nothing with frequency $p_0$. Thus, in weak
pulses QKD, a rather important fraction of the non-empty pulses
actually contain more than one photon. For these pulses, Eve is
then no longer limited by the no-cloning theorem: she can simply
keep some of the photons while letting the others go to Bob. Such
an attack is called {\em photon-number splitting} (PNS) attack.
Although PNS attacks are far beyond today's technology
\cite{felix}, if one includes them in the security analysis, the
consequences are dramatic \cite{brassard,lutkenhaus}.

In this Letter, we present new QKD protocols that are secure
against PNS attack up to significantly longer distances, and that
can thus lead to a secure implementation of QKD with weak pulses.
These protocols are better tailored than the ones studied before
to exploit the correlations that can be established using $\rho$.
The basic idea is that Alice should encode each bit into a pair of
{\em non-orthogonal} states belonging to {\em two or more}
suitable sets.

The structure of the paper is as follows. First, we review the PNS
attack on the first and best-known QKD protocol, the BB84 protocol
\cite{bb84}, in order to understand why this attack is really
devastating when the bit is encoded into pairs of orthogonal
states. Then we present the benefits of using non-orthogonal
states, mostly by focusing on a specific new protocol which is a
simple modification of the BB84.

{\em PNS attacks on the BB84 protocol.} Alice encodes each bit in
a qubit, either as an eigenstate of $\si_x$ ($\ket{+x}$ coding 0
or $\ket{-x}$ coding 1) or as an eigenstate of $\si_z$ ($\ket{+z}$
coding 0 or $\ket{-z}$ coding 1). The qubit is sent to Bob, who
measures either $\si_x$ or $\si_z$. Then comes a classical
procedure known as "sifting" or "basis-reconciliation": Alice
communicates to Bob through a public classical channel the basis,
$x$ or $z$, in which she prepared each qubit. When Bob has used
the same basis for his measurement, he knows that (in the absence
of perturbations, and in particular in the absence of Eve) he has
got the correct result. When Bob has used the wrong basis, the
partners simply discard that item.

Consider now the implementation of the BB84 protocol with weak
pulses. Bob's raw detection rate is the probability that he
detects a photon per pulse sent by Alice. In the absence of Eve,
this is given by \ba R_{raw}(\delta)&=&\sum_{n\geq
1}p_n\,\left(1-(1-\eta_{det}\eta_{\delta})^n\right)\,\simeq\,
\eta_{det}\eta_{\delta}\,\mu\,, \label{raw}\ea where $\eta_{det}$
is the quantum efficiency of the detector (typically 10\% at
telecom wavelengths), and $\eta_{\delta}$ is attenuation due to
the losses in the fiber of length $\ell$: \ba
\eta_{\delta}\,=\,{10}^{-\delta/10}&,\;& \delta=\alpha
\ell\,[\mbox{dB}]\,. \ea Below, when we give a distance, we assume
the typical value $\alpha= 0.25 \mbox{ dB/km}$. The approximate
equality in (\ref{raw}) is valid if
$\eta_{det}\eta_{\delta}\,p_n\,n<<1$ for all $n$, which is always
the case in weak pulses QKD.

If we endow Eve with unlimited technological power within the laws
of physics, the following PNS attack ({\em storage attack}) is in
principle possible \cite{brassard,lutkenhaus}: (I) Eve counts the
number of photons, using a photon-number quantum non-demolition
(QND) measurement; (II) she blocks the single photon pulses, and
for the multi-photon pulses she stores one photon in a quantum
memory; she forwards the remaining photons to Bob using a
perfectly transparent quantum channel, $\eta_{\delta}=1$
\cite{note8}; (III) she waits until Alice and Bob publicly reveal
the used bases and correspondingly measures the photons stored in
her quantum memory: she has to discriminate between two orthogonal
states, and this can be done deterministically. This way, Eve has
obtained full information about Alice's bits, thence no processing
can distill secret keys for the legitimate users; moreover, Eve
hasn't introduced any error on Bob's side.

The unique constraint on PNS attack is that Eve's presence should
not be noticed; in particular, Eve must ensure that the rate of
photons received by Bob (\ref{raw}) is not modified \cite{note2}.
Thus, the PNS attack can be performed on all pulses only when the
losses that Bob expects because of the fiber are equal to those
introduced by Eve's storing and blocking photons, that is, when
the attenuation in the fiber is larger than a critical value
$\delta_c^{BB84}$ defined by \ba R_{raw}(\delta_c^{BB84})&=&
\sum_{n\geq 2}p_n\,\left(1-(1-\eta_{det})^{n-1}\right)\,
\simeq\,\eta_{det}\,p_2\,. \label{pnsbb84}\ea For $\mu=0.1$, we
find ${\delta_c}^{BB84}= 13\mbox{ dB}$, that is
${\ell_c}^{BB84}\approx 50\mbox{ km}$. For shorter distances, Eve
can optimize her attack, but won't be able to obtain full
information; Alice and Bob can therefore use a {\em privacy
amplification} scheme to retrieve a shorter secret key from their
data. In conclusion, for $\delta\geq {\delta_c}^{BB84}$, the
weak-pulses implementation of the BB84 protocol becomes in
principle insecure, even for zero quantum-bit error rate (QBER).

{\em Encoding in non-orthogonal states.} The extreme weakness of
the BB84 protocol against PNS attacks is due to the fact that
whenever Eve can keep one photon, she gets all the information,
because after the sifting phase she has to discriminate between
two eigenstates of a known Hermitian operator. Intuition suggests
then that the robustness against PNS attacks can be increased by
using protocols that encode the classical bit into pairs of
non-orthogonal states, that cannot be discriminated
deterministically. We prove that this intuition is correct.

\begin{center}
\begin{figure}
\epsfxsize=9cm \epsfbox{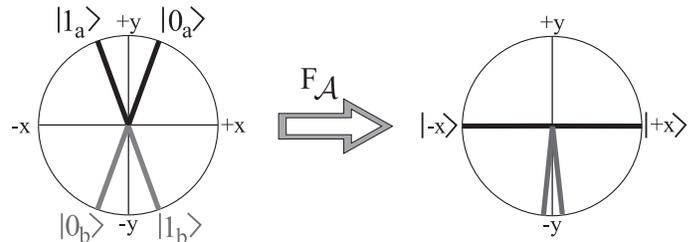} \caption{Two pairs of
non-orthogonal states on the equator of the Poincar\'e sphere, and
the effect of the filter $F_{\cal{A}}$.} \label{figstates}
\end{figure}
\end{center}

To fix the ideas, consider the following protocol using four
states: Alice encodes each bit in the state of a qubit, belonging
either to the set ${\cal{A}}=\big\{\ket{0_a},\ket{1_a}\big\}$ or
to the set ${\cal{B}}=\big\{\ket{0_b},\ket{1_b}\big\}$, with
$|\braket{0_a}{1_a}|=|\braket{0_b}{1_b}|=\chi\neq 0$ (Fig.
\ref{figstates}, left). In the absence of an eavesdropper, Bob can
be perfectly correlated with Alice: in fact, although the two
states are not orthogonal, one can construct a generalized
measurement that unambiguously discriminates between the two. The
price to pay is that sometimes one gets an inconclusive result
\cite{peres}. Such a measurement can be realized by a selective
filtering, that is a filter whose effect is not the same on all
states, followed by a von Neumann measurement on the photons that
pass the filter \cite{bruno}. In the example of Fig.
\ref{figstates}, the filter that discriminates between the
elements of ${\cal{A}}$ is given by
$F_{\cal{A}}=\frac{1}{\sqrt{1+\chi}}\big(\ket{+x}\bra{{1_a}^{\perp}}
+\ket{-x}\bra{{0_a}^{\perp}}\big)$, where $\ket{\psi^{\perp}}$ is
the state orthogonal to $\ket{\psi}$. When the photons are
prepared in a state of the pair ${\cal{A}}$, a fraction $1-\chi$
of them pass this filter, and in this case the von-Neumann
measurement of $\sigma_x$ achieves the discrimination. It is then
clear how the cryptography protocol generalizes BB84: Bob randomly
applies on each qubit one of the two filters $F_{\cal{A}}$ or
$F_{\cal{B}}$, and measures $\si_x$ on the outcome. Later, Alice
discloses for each bit the set ${\cal{A}}$ or ${\cal{B}}$: Alice
and Bob discard all the items in which Bob has chosen the wrong
filter and all the inconclusive results.

Of course, since not all the qubits will pass the filter even when
it was correctly chosen, there is a small nuisance on Bob's side
because the net key rate is decreased. This is compensated by
increasing $\mu$ by a factor $1/(1-\chi)$. However, the nuisance
is by far bigger on Eve's side, even when the increased mean
number of photons $\mu$ is taken into account. We shall give a
detailed analysis of the PNS attacks below for a specific
protocol, but a simple estimate shows the origin of the improved
robustness. Eve can obtain {\em full information} only when (i)
she can block all the pulses containing one and two photons, and
(ii) on the pulses containing three or more photons, she performs
a suitable unambiguous discrimination measurement (see below) and
obtains a conclusive outcome, which happens only with probability
$p_{ok}<1$. Consequently, the critical attenuation is defined by
$R_{raw}(\delta_c)\simeq \eta_{det}\,p_3
(\frac{\mu}{1-\chi})\,p_{ok}$, and is determined by $p_3$ instead
of $p_2$ as in the BB84, see (\ref{pnsbb84}). For typical values,
$\delta_c-\delta_c^{BB84}\approx 10$ dB, which means an
improvement of some 40km in the distance \cite{note9}.

{\em A specific protocol.} Here is an astonishingly simple
protocol using four non-orthogonal states. Alice sends randomly
one of the four states $\ket{\pm x}$ or $\ket{\pm z}$; Bob
measures either $\si_x$ or $\si_z$. Thus, at the "quantum" level,
the protocol is identical to BB84, and can be immediately
implemented with the existing devices. However, we {\em modify the
classical sifting procedure}: instead of revealing the basis,
Alice announces publicly one of the four pairs of non-orthogonal
states ${\cal{A}}_{\omega,\omega'} =\big\{\ket{\omega x
},\ket{\omega' z}\big\}$, with $\omega,\omega'\in\{+,-\}$, and
with the convention that $\ket{\pm x}$ code for 0 and $\ket{\pm
z}$ code for 1. Within each set, the overlap of the two states is
$\chi=\frac{1}{\sqrt{2}}$. Because of the peculiar choice of
states, the usual procedure of choosing randomly between $\si_x$
or $\si_z$ turns out to implement the most effective unambiguous
discrimination. For definiteness, suppose that for a given qubit
Alice has sent $\ket{+x}$, and that she has announced the set
${\cal{A}}_{+,+}$. If Bob has measured $\si_x$, which happens with
probability $\demi$, he has certainly got the result $+1$; but
since this result is possible for both states in the set
${\cal{A}}_{+,+}$, he has to discard it. If Bob has measured
$\si_z$ and got $+1$, again he cannot discriminate. But if he has
measured $\si_z$ and got $-1$, then he knows that Alice has sent
$\ket{+x}$ and adds a 0 to his key. By symmetry, we see that after
this sifting procedure Bob is left with $\frac{1}{4}$ of the raw
list of bits, compared to the $\frac{1}{2}$ of the original BB84
protocol. Thus, for a fair comparison with BB84 using $\mu=0.1$,
we shall take here $\mu=0.2$, so that the net key rates without
eavesdropper at a given distance are the same for both protocols.
In spite of the fact that a larger $\mu$ is used (that is,
multi-photon pulses are more frequent), this new protocol is
provably better than BB84 against PNS attacks at QBER$=0$. This is
our main claim, and is demonstrated in the following.

{\em PNS attacks at QBER=0.} First, let us prove something that we
mentioned above, namely: for protocols using four states like the
one under study, Eve can obtain full information from three-photon
pulses by using strategies based on unambiguous
state-discrimination. Such strategies have also been considered
for BB84, because (although worse than the storage attack for an
all-powerful Eve) they don't require a quantum memory
\cite{dusek}, and in their simplest implementation the
photon-number QND measurement is not required either \cite{yuen}.
The most powerful of these attacks, against which any protocol
using four states becomes completely insecure for the three-photon
pulses, goes as follows \cite{lutkpriv}. A pulse containing three
photons is necessarily in one of the four states
$\ket{\Psi_1}=\ket{+z}^{\otimes 3}$, $\ket{\Psi_2}=
\ket{+x}^{\otimes 3}$, $\ket{\Psi_3}=\ket{-z}^{\otimes 3}$,
$\ket{\Psi_4}= \ket{-x}^{\otimes 3}$; that is, in the {\em
symmetric} subspace of 3 qubits. The dimension of this subspace is
4, and it can be shown that all the $\ket{\Psi_k}^{\otimes {3}}$
are linearly independent \cite{theorem}. Therefore, there exist a
measurement $\cal{M}$ that distinguishes unambiguously among them,
with some probability of success. In the present case, there exist
even four orthogonal states of three qubits, $\ket{\Phi_k}$,
$k=1,...,4$, such that $|\braket{\Phi_i}{\Psi_j}|
=\frac{1}{\sqrt{2}}\,\delta_{ij}$ \cite{note4}. The measurement
$\cal{M}$ is then any von-Neumann measurement discriminating the
$\ket{\Phi_k}$; it will give a conclusive outcome with probability
$p_{ok}=\frac{1}{2}$, which is optimal \cite{theorem,toni}.

It is then clear that Eve can obtain full information if she can
block all the one- and two-photons pulses and half of the
three-photon pulses, by applying the following PNS attack: (I) she
measures the number of photons; (II) she discards all pulses
containing less than 3 photons; (III) on the pulses containing at
least 3 photons, she performs $\cal{M}$, and if the result is
conclusive (which happens with probability $p_{ok}\gtrsim \demi$)
she sends a new photon prepared in the good state to Bob. We refer
to this attack as to {\em intercept-resend with unambiguous
discrimination (IRUD) attack}. Neither the quantum memory is
needed, nor is the lossless channel, since the new state can be
prepared by a friend of Eve located close to Bob.

The critical attenuation $\delta_c$ at which the IRUD attack
becomes always possible is defined by $\eta_{\delta_c}\mu=
p_{ok}p_3(\mu)$; for $\mu=0.2$, this gives $\delta_c =
25.6\,\mbox{dB}\approx 2 \delta_c^{BB84}$. Thus, the ultimate
limit of robustness (in the case of zero errors) is shifted from
$\sim 50$km up to $\sim 100$km by using our simple modification of
the BB84 protocol. To further increase the limit of 100km, one can
move to protocols using six or more non-orthogonal states
\cite{toni}.

\begin{center}
\begin{figure}
\epsfxsize=7cm \epsfbox{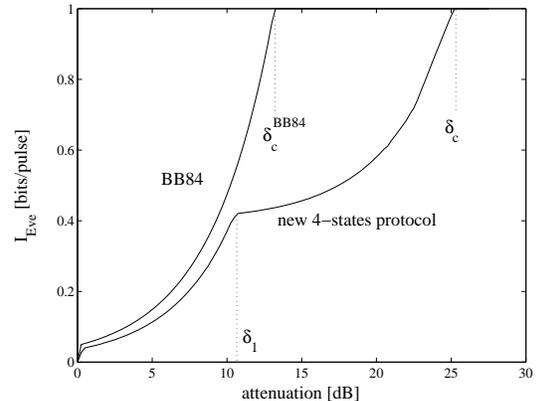} \caption{PNS attacks with
QBER=0 on the BB84 protocol for $\mu=0.1$ and on the new protocol
for $\mu=0.2$: Eve's information as a function of the attenuation
$\delta=\alpha\ell$.} \label{qberzero}
\end{figure}
\end{center}

Figure \ref{qberzero} plots Eve's information for the best PNS
attack at QBER$=0$, as a function of the attenuation. Note that
the new protocol is better than BB84 at any distance. For almost
all $\delta<\delta_c$, the best PNS attack is not the IRUD but a
storage attack, in which Eve keeps one or two photons in a quantum
memory and waits for the announcements of the sifting phase.
Recall that in BB84, this kind of attack provides Eve with full
information. In our protocol Alice announces sets of two
non-orthogonal states, so storage attacks give Eve only a limited
amount of information. If Eve keeps $n$ photons and the overlap is
$\chi$ (here, $1/\sqrt{2}$), the largest information she can
obtain is $I(n,\chi)=1-H(P,1-P)$ with
$P=\demi(1+\sqrt{1-\chi^{2n}})$ \cite{peres}. In particular, Eve
obtains $I(1,\frac{1}{\sqrt{2}})\approx 0.4$ bits/pulse for the
attenuation $\delta_1$ at which she can always keep one photon
($\delta_1\simeq 11$ dB for $\mu=0.2$).

In conclusion: in the limiting case of QBER$=0$, our protocol is
always more secure than BB84 against PNS attacks, and can be made
provably secure against such attacks in regions where BB84 is
already provably insecure. Recall that the comparison is made by
fixing the net key rates without eavesdropper at a given distance.

{\em Attacks at QBER$>$0 on the new protocol.} In real
experiments, dark counts in the detectors and misalignement of
optical elements always introduce some errors. It is then
important to show that the specific protocol we presented does not
break down if a small amount of error on Bob's side is allowed.
Several attacks at non-zero QBER are described in detail in Ref.
\cite{toni}. Here, we sketch the analysis of two individual
attacks.

First, let us suppose that Eve uses the {\em phase-covariant
cloning machine} that is the optimal individual attack against
BB84 \cite{cloners}. In the case of the present protocol, Eve can
extract less information from her clones, again because Alice does
not disclose a basis but a set of non-orthogonal states. As a
consequence, the condition $I_{Bob}=I_{Eve}$ is fulfilled up to
QBER=15\% \cite{toni}, a value which is slightly higher than the
14,67\% obtained for BB84. So our new protocol, designed to avoid
PNS attacks in a weak-pulses implementation, seems to be robust
also against individual eavesdropping in a single-photon
implementation. Incidentally note that, in the case of a
single-photon implementation, our protocol is at least as secure
as the B92 protocol in the sens of "unconditional security" proofs
\cite{uncond}. This is because our protocol can be seen as a
modified B92, where Alice chooses randomly between four sets of
non-orthogonal states \cite{sofyan}.

The second kind of individual attacks that we like to discuss, and
that we call {\em PNS+cloning attacks}, are specific to imperfect
sources. Focus on the range $\delta\simeq 10 -20\mbox{ dB}$ (see
Fig. \ref{qberzero}), where one-photon pulses can be blocked and
the occurrence of three or more photons is still comparatively
rare. Because for the BB84 Eve has already full information in
this range, such attacks have never been considered before. Eve
could take the two photons, apply an asymmetric $2\rightarrow 3$
cloning machine and send one of the clones to Bob; she keeps two
clones and some information in the machine. By a suitable choice
of the cloning machine, the QBER at which $I_{Bob}=I_{Eve}$ is
lowered down to $\sim 9$\% \cite{toni}. In Ref. \cite{damien}, a
successful qubit distribution over 67km with $\mu=0.2$ and
QBER$=5\%$ has been reported. Under the considered PNS attacks,
such distribution is provably insecure using the sifting procedure
of BB84, while it can yield a secret key if our sifting procedure
is used.

In summary, we have shown that by encoding a classical bit in sets
of non-orthogonal qubit states, quantum cryptography can be made
significantly more robust against photon-number splitting attacks.
We have presented a specific protocol, which is identical to the
BB84 protocol for all the manipulations at the quantum level and
differs only in the classical sifting procedure. Under the studied
attacks, our protocol is secure in a region where BB84 is provably
insecure. Preliminary studies of more complex attacks suggest that
it is at least as robust as BB84 in any situation, and could then
replace it. Moreover, our encoding can easily be combined with
more complex procedures on the quantum level, e.g. \cite{hwang}.

We thank Norbert L\"utkenhaus and Daniel Collins for insightful
comments. We acknowledge financial supports by the Swiss OFES and
NSF within the European IST project EQUIP and the NCCR "Quantum
Photonics".

\end{multicols}


\begin{thebibliography}{99}
\bibitem{note2} In a realistic setup, Bob can monitor only the
detection rate and the fraction of double counts in the wrong
basis.
\bibitem{note9} If we'd allow Eve an access to Bob's detectors,
both critical attenuations become smaller, but the improvement
$\delta_c-\delta_c^{BB84}\approx 10$ dB holds unchanged, because
it is independent of $\eta_{det}$.
\end{thebibliography}
\end{document}